\documentclass[12pt,preprint]{aastex}

\shorttitle{
	Accelerating universe from gravitational leakage into extra dimensions:
	confrontation with SNeIa
	}	
\shortauthors{Zhu, Z.-H. \& Alcaniz, J. S.}

\begin{document}

\title{
	Accelerating universe from gravitational leakage into extra dimensions:
		confrontation with SNeIa
	}

\author{Zong-Hong Zhu}
\affil{Department of Astronomy, Beijing Normal University,
		Beijing 100875, China}
\email{zhuzh@bnu.edu.cn}

	\and

\author{Jailson S. Alcaniz}
\affil{Departamento de Astronomia, Observat\'orio Nacional,
		20921-400 Rio de Janeiro - RJ, Brasil}
\email{alcaniz@on.br}

\begin{abstract}
There is mounting observational evidence that the expansion of our universe 
  is undergoing an acceleration.
A dark energy component has usually been invoked as the most feasible 
  mechanism for the acceleration.
However, it is desirable to explore alternative possibilities motivated by
  particle physics before adopting such an untested entity.
In this work, we focus our attention on an acceleration mechanism:
  one arising from gravitational leakage into extra dimensions.
We confront this scenario with high-$z$ type Ia supernovae compiled by
  Tonry et al. (2003) and recent measurements of the X-ray gas mass 
  fractions in clusters of galaxies published by Allen et al. (2002,2003).
A combination of the two databases gives at a 99\% confidence level that
  $\Omega_m=0.29^{+0.04}_{-0.02}$, $\Omega_{rc}=0.21^{+0.08}_{-0.08}$, and
  $\Omega_k=-0.36^{+0.31}_{-0.35}$, indicating a closed universe.
We then constrain the model using the test of the turnaround redshift,
  $z_{q=0}$, at which the universe switches from deceleration to acceleration. 
We show that, in order to explain that acceleration happened earlier than 
  $z_{q=0} = 0.6$ 
  within the
  framework of gravitational leakage into extra dimensions, a low matter 
  density, $\Omega_m < 0.27$, or a closed universe is necessary.
\end{abstract}


%

\section{Introduction}

The recent well known distance measurements of distant type Ia supernovae
  (SNeIa) suggest an accelerating universe at large scales
  (Riess et al. 1998, Perlmutter et al. 1999, Tonry et al. 2003,
	Barris et al. 2004, Knop et al. 2003, Riess et al. 2004).
The cosmic acceleration has also been confirmed, independently of the SNeIa
  magnitude-redshift relation, by the observations of the cosmic microwave
  background anisotropies (WMAP: Bennett et al. 2003)
  and the large scale structure in the distribution of galaxies (SDSS:
  Tegmark et al. 2003a,b).
It is well known that all known types of matter with positive pressure
  generate attractive forces and decelerate the expansion of the universe.
Given this, a dark energy component with negative pressure was generally 
  suggested
  to be the invisible fuel that drives the current acceleration of the 
  universe. There are a huge number of candidates for the dark energy 
  component in the literature, such as
  a cosmological constant $\Lambda$ 
	(Carroll et al. 1992; 
	Krauss and Turner 1995; 
	Ostriker and Steinhardt 1995;
	Chiba and Yoshii 1999),
  a decaying vacuum energy density or a time varying $\Lambda$-term
	(Ozer and Taha 1987; Vishwakarma 2001),
  an evolving scalar field 
	(referred to by some as quintessence: 
	Ratra and Peebles 1988; 
	Caldwell et al. 1998;
	Wang and Lovelace 2001;
	Weller and Albrech 2002;
	Gong 2002;
	Li et al. 2002a,b;
	Chen and Ratra 2003;
	Mukherjee et al. 2003;
	Gong 2004),
  the phantom energy, in which the sum of the pressure and energy
    density is negative
        (Caldwell 2002;
        Dabrowski et al. 2003;
	Wang, Gong and Su 2004),
  the so-called ``X-matter" 
	(Turner and White 1997; 
	Zhu 1998;
	Podariu and Ratra 2001;
	Zhu, Fujimoto and Tatsumi 2001; 
	Alcaniz, Lima and Cunha 2003; 
	Lima, Cunha and Alcaniz  2003;
	Feng, Wang and Zhang 2004;
	Dai, Liang and Xu 2004),
  the Chaplygin gas 
	(Kamenshchik et al. 2001; 
	Bento et al. 2002; 
	Alam et al. 2003; 
	Alcaniz, Jain and Dev 2003; 
	Dev, Alcaniz and Jain 2003; 
	Silva and Bertolami 2003;
	Makler et al. 2003),
  and the Cardassion model
	(Freese and Lewis 2002;
	Zhu and Fujimoto 2002, 2003;
	Sen and Sen 2003;
	Wang et al. 2003;
	Frith 2004;
	Gong and Duan 2004a,b).

However, the dark energy has so far no convincing direct laboratory evidence for
  its existence, so it is desirable to explore alternative 
  possibilities motivated by particle physics before adopting such a component.
In this respect the models that make use of the very ideas of
  branes and extra dimensions to obtain an accelerating universe 
  are particularly interesting
	(Randall and Sundrum 1999a,b).
Within the framework of these braneworld cosmologies, our observable universe
  is assumed to be a surface or a brane embedded in a higher dimensional bulk
  spacetime in which gravity could spread, and the bulk gravity sees its own 
  curvature term on the brane which accelerates the universe without dark 
  energy (Randall 2002). 
Recently, based on the model of Dvali et al. (2000) of brane-induced gravity,
  Deffayet and coworkers (Deffayet 2001, Deffayet, Dvali and Gabadadze 2002) 
  proposed a
  scenario in which the observed late time acceleration of the expansion of
  the universe is caused by gravitational leakage into an extra dimension
  and the Friedmann equation is modified as follows
\begin{equation}
\label{eq:ansatz}
H^2 = H_0^2 \left[ 
	\Omega_k(1+z)^2+\left(\sqrt{\Omega_{rc}}+ 
	\sqrt{\Omega_{rc}+\Omega_m (1+z)^3}\right)^2
		\right]
\end{equation}
where $H$ is the Hubble parameter as a function of redshift $z$ ($H_0$ is its
  value at the present), $\Omega_k$, $\Omega_{rc}$ and $\Omega_m$ represent
  the fractional contribution of curvature, the bulk-induced term and the 
  matter (both baryonic and nonbaryonic), respectively. 
$\Omega_{rc}$ is defined as $\Omega_{rc} \equiv 1/4r_c^2H_0^2$, where $r_c$
  is the crossover scale beyond which the gravitational force follows the 
  5-dimensional $1/r^3$ behavior.
From a phenomenological standpoint, it is a testable scenario with the same
  number of parameters as a cosmological constant model, contrasting with 
  models of quintessence that have an additional free function, the equation
  of state, to be determined (Deffayet et al. 2002).
Such a possible mechanism for cosmic acceleration has triggered investigations 
  aiming to constrain this scenario using various cosmological observations,
  such as 
  SNeIa (Avelino and Martins 2002; Deffayet, Dvali and Gabadadze 2002; 
	 Deffayet et al. 2002; Dabrowski et al. 2004),
  angular size of compact radio sources (Alcaniz 2002),
  the age measurements of high-$z$ objects (Alcaniz, Jain and Dev 2002),
  the optical gravitational lensing surveys (Jain et al. 2002) and
  the large scale structures (Multam\"aki et al. 2003).
But the results are disperse and somewhat controversial, with most of them
  claiming good agreement between data and the model while some of them ruling
  out gravitational leakage into an extra dimension as a feasible mechanism
  for cosmic acceleration.

The purpose of this work is to quantitatively confront the scenario with
  the updated SNeIa sample compiled by Tonry et al. (2003) and to try to 
  constrain the model parameters more accurately.
It is shown that, although the two parameters, $\Omega_{rc}$ and $\Omega_m$, 
  are degenerate and there is a range on the parameter plane to be consistent
  with the SNeIa data, a closed universe is prefered by this scenario.
As is well known, the measurement of the X-ray gas mass fraction in galaxy
  clusters is an efficient way to determine the matter density, $\Omega_m$,
  and hence can be used for breaking the degeneracy between 
  $\Omega_{rc}$ and $\Omega_m$.
When we combine the X-ray database published by Allen et al. (2002, 2003) for
  analyzing, we obtained a closed universe 
  at a 99\% confidence level, i.e., for the scenario of gravitational
  leakage into an extra dimension, a universe with curvature is favored by
  the combination of the two databases.
We also analyze the turnaround redshift, $z_{q=0}$, at
  which the universe switches from deceleration to acceleration within the
  framework of the scenario.
It is shown that, if the turnaround redshift happened earlier than 
  $z_{q=0} = 0.6$, only a low matter density, $\Omega_m < 0.27$, or a closed
  universe can explain this transition epoch.
If, however, we consider the recent estimate by Riess et al. (2004), i.e.,
  $z_{q=0} = 0.46 \pm 0.13$, then a spatially flat scenario with $\Omega_m =0.3$
  (as suggested by clustering estimates) predicts $z_{q=0} = 0.48$, which is
  surprisingly close to the central value given by Riess et al. (2004). 
The paper is organized as follows.
In the next section, we consider the observational constraints on the parameter
  space of the scenario arising from the updated SNeIa sample compiled by
  Tonry et al. (2003), as well as the combination with the X-ray gas mass
  fractions in galaxy clusters published by Allen et al. (2002, 2003).
In section~3 we discuss the bounds on the model from the turnaround redshift,
  $z_{q=0}$.
Finally, we present our conclusion and discussion in section~4.


\section{Constraints from SNeIa and galaxy cluster data}

Recently, Tonry et al. (2003) compiled a large database of SNeIa
  from the literature and eight new SNeIa from the High-$z$ Supernova
  Search Team.
Since the techniques for data analysis vary between individual SNeIa samples,
  the authors have attempted to recompute the extinction estimates and the
  distance determination through the MLCS fitting (Riess et al. 1998), 
  the $\Delta m_{15}$ method of Phillips et al. (1999), the modified dm15 
  fitting (Germany 2001) and the BATM method (Tonry 2003).
Zero-point differences between each method were computed by comparing common
  SN measurements, and distances were placed on a Hubble flow zeropoint 
  ($dH_0$), and the median selected as the best distance estimate (for more
  details of this procedure, see Tonry et al. 2003; Barris et al. 2004).
Tonry et al. (2003) present redshift and distance for 230 SNeIa, which
  includes many objects unsuitable for cosmological analysis, such as the SNeIa
  being heavily extinguished or nearby enough for velocity uncertainties to be
  a major problem.
To determine cosmological parameters, the authors used  a redshift cut of
  $z > 0.01$ and an extinction cut of $A_V < 0.5$ mag.
The resulting sample of 172 SNeIa is illustrated on a residual Hubble Diagram
  with respect to an empty universe ($\Omega_m = 0$, $\Omega_{rc} = 0$) 
  in Figure~1.
We will use this sample to give an observational constraint on the 
  model parameters, $\Omega_{rc}$ and $\Omega_m$.

%
%

For the ansatz (1), we are required to calculate $\chi^2$ as a function of the 
  model parameters ($\Omega_m$, $\Omega_{rc}$) and the Hubble constant $H_0$. 
Following Tonry et al. (2003), we added 500 km s$^{-1}$ divided by the redshift
  in quadrature to the distance error given in their Table~15 for calculating
  $\chi^2$.
In order to concentrate solely on the density parameters, we need to 
  marginalize over the Hubble constant $H_0$.
Since $H_0$ appears as a quadratic term in $\chi^2$, it appears as a separable
  Gaussian factor in the probability to be marginalized over.
Thus marginalizing over $H_0$ is equivalent to evaluating $\chi^2$ at its
  minimum with respect to $H_0$ (Barris et al. 2004). 
This procedure allows us to determine contours of constant probability density
  for the model parameters ($\Omega_m$, $\Omega_{rc}$) corresponding to 68\%,
  95\%, and 99\% confidence levels, which is shown in Figure~2.
The best fit happens at $\Omega_m = 0.43$ and $\Omega_{rc} = 0.26$.
As is shown in Figure~2, although there is a range on the parameter plane to 
  be consistent with the SNeIa data, a closed universe is favored.
Furthmore, the two density parameters, $\Omega_{rc}$ and $\Omega_m$, are
  highly degenerate, which is very similar to the degeneracy between
  $\Omega_{\Lambda}$ and $\Omega_m$ found by Tonry et al. (2003).
In order to determine $\Omega_{rc}$ and $\Omega_m$ respectively, an independent
  measurement of $\Omega_{rc}$ or $\Omega_m$ is needed.
As shown below, the X-ray gas mass fraction data of galaxy clusters are 
  appropriate for this purpose, because the data are only sensitive to 
  $\Omega_m$ (Allen et al. 2002, 2003).

%
%

Since clusters of galaxies are the largest virialized systems in the universe,
  their matter content should provide a fair sample of the matter content of
  the universe as a whole, and a comparison of their gas mass fractions,
  $f_{\rm gas} = M_{\rm gas} / M_{\rm tot}$,
  as inferred from X-ray observations, with the cosmic
  baryon fraction can provide a direct constraint on the density parameter
  of the universe $\Omega_m$ (White et. al. 1993).
Moreover, assuming the gas mass fraction is constant in cosmic time,
  Sasaki (1996) shows that the $f_{\rm gas}$ measurements of clusters of 
  galaxies at different redshifts also provide a way to constrain other
  cosmological parameters describing the geometry of the universe.
Recently, Allen et al. (2002; 2003) published the $f_{\rm gas}$ profiles for 
  the 10 relaxed clusters observed by the {\it Chandra} satellite.
Except for Abell 963, the $f_{\rm gas}$ profiles of the other 9 clusters
  appear to have converged or be close to converging with $r_{2500}$, 
  the radius within which the mean mass density is 2500 times the critical 
  density of the universe at the redshift of the cluster.
The gas mass fraction values of these 9 clusters were shown in Figure~5
  of Allen et al. (2003).
This database can be used to break the degeneracy between $\Omega_{rc}$ and 
  $\Omega_m$ mentioned above, since it has been shown that the X-ray gas
  mass fraction is mostly sensitive to $\Omega_m$ no matter what the 
  cosmological model is (Allen et al. 2002; Lima et al. 2003). 
The probability density over the model parameters, $\Omega_{rc}$ and 
  $\Omega_m$, for the 9 galaxy clusters is calculated using the method
  described in Allen et al. (2002).
Following Allen et al. (2003), we include Gaussian priors 
  on the bias factor, $b = 0.93 \pm 0.05$, a value appropriate for hot 
  ($kT > 5$ KeV) clusters from the simulations of Bialek et al. (2001), 
  on the Hubble constant, $h = 0.72 \pm 0.08$, the final result from the 
  Hubble Key Project by Freedman et al. (2001),
  and on $\Omega_m h^{2} = 0.0205 \pm 0.0018$ (O'Meara et al. 2001), from
  cosmic nucleosynthesis calculations constrained by the observed abundances
  of light elements at high redshifts.
We then multiply the probability densities from the 172 SNeIa and the 9
  galaxy clusters, and obtain our final results on $\Omega_{rc}$ and
  $\Omega_m$, which are shown in Figure~3.
 
Figure~3 illustrates the 68\%, 95\% and 99\% confidence levels in the
  ($\Omega_m$,$\Omega_{rc}$) plane using the red, green and yellow shaded
  areas, respectively.
Our fits give at a 99\% confidence level that
  $\Omega_m=0.29^{+0.04}_{-0.02}$, $\Omega_{rc}=0.21^{+0.08}_{-0.08}$, and
  $\Omega_k=-0.36^{+0.31}_{-0.35}$.
Although there is a range on the parameter plane being consistent with both
  the SNeIa and galaxy clusters data, and the resulting matter density
  $\Omega_m$ is reasonable, a closed universe is obtained at a 99\%
  confidence level, which is inconsistent with the result, 
  $\Omega_k=-0.02^{+0.02}_{-0.02}$, found by the
  WMAP (Bennett et al. 2003).
Avelino and Martins (2002) analyzed the same model with the 92
  SNeIa from Riess et al. (1998) and Perlmutter et al. (1999).
Assuming a flat universe, the authors obtained a very low matter
  density and claimed the model was disfavorable.
In additional to including new SNeIa data from Tonry et al. (2003), 
  and combining the X-ray data of 9 galaxy clusters, 
  we relax the flat universe constraint in their analysis. 
We obtained a reasonable matter density, but a closed universe.
In some sense, i.e., if we assume that our Universe is spatially flat, 
  as indicated by WMAP results, the accelerating scenario from 
  gravitational leakage into extra dimension does not seem to be favored 
  by observational data. 
However, two points should be emphasized here. 
  First, that the same conclusion happens by performing a similar analysis 
  with our current standard model, i.e., a $\Lambda$CDM universe. Second, 
  that we have made heavy use of the X-ray gas mass fraction in
  clusters to determine the matter density.
This kind of analysis depends on the assumption that $f_{\rm gas}$ values
  should be invariant with redshift, which has been criticised by a minority
  of works in the field.
For example, a recent comparison of distant clusters observed by XMM-Newton
  and Chandra satellites with available local cluster samples indicate a
  possible evolution of the $M$--$T$ relation with redshift, i.e., the standard
  paradigm on cluster gas physics need to be revised (Vauclair et al. 2003).
We should keep this point in mind when we make the conclusion that the
  gravitational leakage scenario is disfavored by the databases.

%
%


\section{Constraints from the turnaround redshift from deceleration
	 to acceleration}

Since the scenario of gravitational leakage into extra dimensions is 
  proposed as a possible mechanism for the cosmic acceleration,
  the turnaround redshift from deceleration to acceleration is expected
  to provide an efficient way for verifying the model.
It can be shown that the deceleration parameter as a function of redshift as
  well as the model parameters takes the form (Zhu and Fujimoto 2004)
\begin{equation}
\label{eq:deceleration}
q(z) \equiv -{\ddot{R}R\over {\dot{R}}^2}
     = -1 + {1 \over 2}{{\rm d}\ln E^2 \over {\rm d}\ln (1+z)}
\end{equation}
where $E^2(z; \Omega_{rc}, \Omega_m) = H^2(z; \Omega_{rc}, \Omega_m) / H_0$.
From Eq.(1), we could derive the turnaround redshift at which
  the universe switches from deceleration to acceleration, or in other words
  the redshift at which the deceleration parameter vanishes, which is as
  follows
\begin{equation}
\label{eq:zq=0}
(1+z)_{q=0} = 2\left( {\Omega_{rc} \over \Omega_m} \right)^{1/3}
\end{equation}
We have shown that Eq. (3) is generally valid no matter what the curvature of
  the universe is, though it was first obtained by Avelino and Martins (2002)
  for a flat universe.
According to Turner and Riess (2002), the value for the turnaround redshift
  lies in the $1\sigma$ interval $0.6 < z_{q=0} <1.7$.
In Fig.4, the two dashed lines represent $z_{q=0} = 0.6$ and 
  $z_{q=0} = 1.7$, respectively, while the hatched region at lower right
  corresponds to $z_{q=0} \le 0$, which means a decelerating universe.
The thick solid line is the flat universe.
The vertical strip with cross-hatching is the matter density 
  $\Omega_m=0.330\pm 0.035$ found by Turner (2002), and the
  vertical dot-dashed lines are $\Omega_m=0.2, 0.4$, a wider range. 
As is shown, in order to explain that cosmic acceleration
  started earlier than $z_{q=0} = 0.6$, either a low matter density,
  $\Omega_m < 0.27$, is needed on the assumption of a flat universe, 
  or a closed universe is necessary for a higher matter density.
If, however, we consider the recent estimate by Riess et al. (2004), i.e.,
  $z_{q=0} = 0.46 \pm 0.13$, then a spatially flat scenario with $\Omega_m=0.3$  (as suggested by clustering estimates) predicts $z_{q=0} = 0.48$, which is
  surprisingly close to the central value given by Riess et al. (2004).



\section{Conclusion and discussion}

The mounting observational evidences for an accelerating universe have
  stimulated renewed interest for alternative cosmologies.
Generally, a dark energy component with negative pressure is invoked to explain
  the SNeIa results and to reconcile the inflationary flatness prediction
  ($\Omega_T = 1$) with the dynamical estimates of the quantity of matter
  in the universe ($\Omega_m \sim 0.3$). 
In this paper we have focused our attention on another possible acceleration
  mechanism, one arising from gravitational leakage into extra dimensions.
In order to be consistent with the current SNeIa and the X-ray clusters data, 
  one would need a closed universe.

Recently Lue et al. (2004) derived dynamical equations for spherical
  perturbations at subhorizon scales and computed the growth of large-scale
  structure in the framework of this scenario.
A suppression of the growth of density and velocity perturbations was
  found, e.g., for $\Omega_m=0.3$, a perturbation of
  $\delta_i=3\times 10^{-3}$ at $z_i=1000$ collapse in the $\Lambda$CDM
  case at $z\approx 0.66$ when its linearly extrapolated density contrast
  is $\delta_c=1.689$, while for the model being considered the collapse
  happens much later at $z\approx 0.35$ when its $\delta_c=1.656$.
Furthermore, the authors showed that this scenario for cosmic acceleration
  gave rise to a present day fluctuation power spectrum normalization
  $\sigma_8 \leq 0.8$ at a 2$\sigma$ level, lower than observed value    
  (Lue et al. 2004).

As is shown in Figure~2 of Deffayet, Dvali and Gabadadze (2002), on the 
  assumption of a flat universe, luminosity distance for $\Lambda$CDM increases
  with redshift faster than that for the model being considered does (for the
  same $\Omega_m$).
Therefore it is natural that, if the $\Lambda$CDM model with 
  ($\Omega_m=0.3$, $\Omega_{\Lambda}=0.7$, $\Omega_k=0$) is consistent with
  the SNeIa data, the gravitational leakage model with 
  ($\Omega_m=0.3$, $\Omega_{rc}=0.1225$, $\Omega_k=0$) will not be as the data
  are becoming enough to determine the cosmological parameters more precisely.
While Deffayet et al. (2002) showed that the gravitational leakage scenario
  was consistent with the 54 SNeIa of the sample C from Perlmutter et al.
  (1999) -- see also Alcaniz \& Pires (2004) -- Avelino and Martins (2002) 
  claimed that this proposal was disfavored by the dataset of 92 SNeIa from 
  Riess et al (1998) and Perlmutter et al. (1999) [combining them via the 
  procedure described in Wang (2000) and Wang \& Garnavich (2001)].
We, however, think that only with a more general analysis, a joint 
  investigation involving different classes of cosmological tests, it will be
  possible to delimit the $\Omega_{\rm{m}} - \Omega_{r_c}$ plane more precisely,
  as well as to test more properly the consistency of these senarios.
Such an analysis will appear in a forthcoming communication 
  (Alcaniz \& Zhu 2004).

\acknowledgements

We would like to thank 
  John. L. Tonry for important clarifications in marginalizing over the Hubble
    constant,
  S. Allen for sending me their compilation of the X-ray mass fraction data and
  W. Li for helpful discussions.
Our thanks go to the anonymouse referee and the editor, Prof. Ethan Vishniac, 
  for valuable comments and useful 
  suggestions, which improved this work very much.
Z.-H. Zhu acknowledges support from the National Natural Science Foundation
  of China 
	and
  the National Major Basic Research Project of China (G2000077602).
J. S. Alcaniz is partially supported by CNPq (305205/02-1) and CNPq 
  (62.0053/01-1-PADCT III/Milenio).

\clearpage

\begin{figure}
\plotone{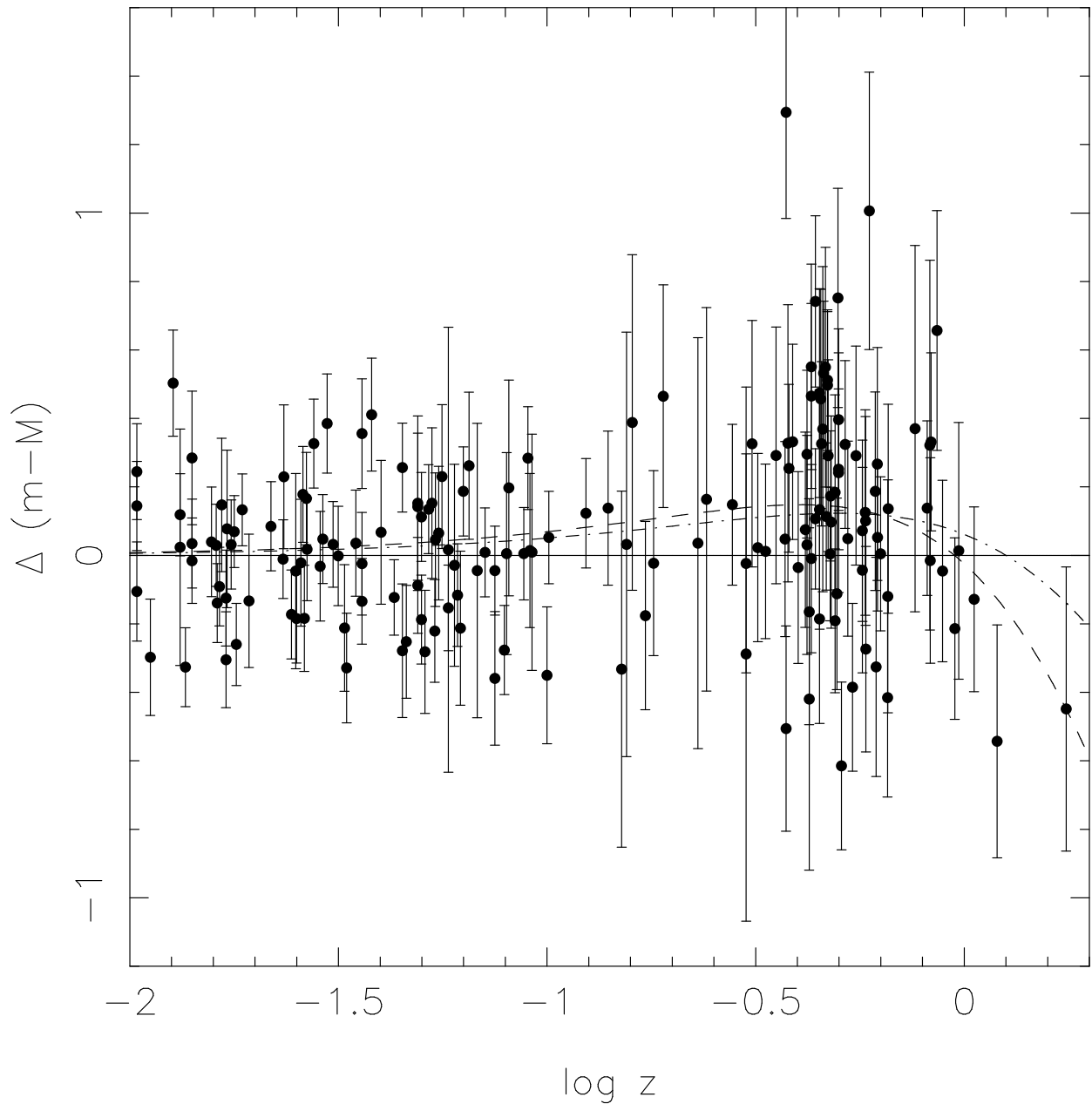}
\figcaption{The 172 SNeIa data points, obtained by imposing constraints 
	    $A_V < 0.5$ and $z > 0.01$ on the 230 SNeIa sample of 
	    Tonry et al. (2003), are shown in a residual Hubble diagram 
	    with respect to an empty universe.
	    The dashed and dot-dashed lines show
	      ($\Omega_m$, $\Omega_{r_c}$) = (0.43, 0.26), our best fit, and 
	      ($\Omega_m$, $\Omega_{\Lambda}$) = (0.3, 0.7), the standard
	      $\Lambda$CDM model, respectively.
	    \label{Fig_data}
           }
\end{figure}

\clearpage

\begin{figure}
\plotone{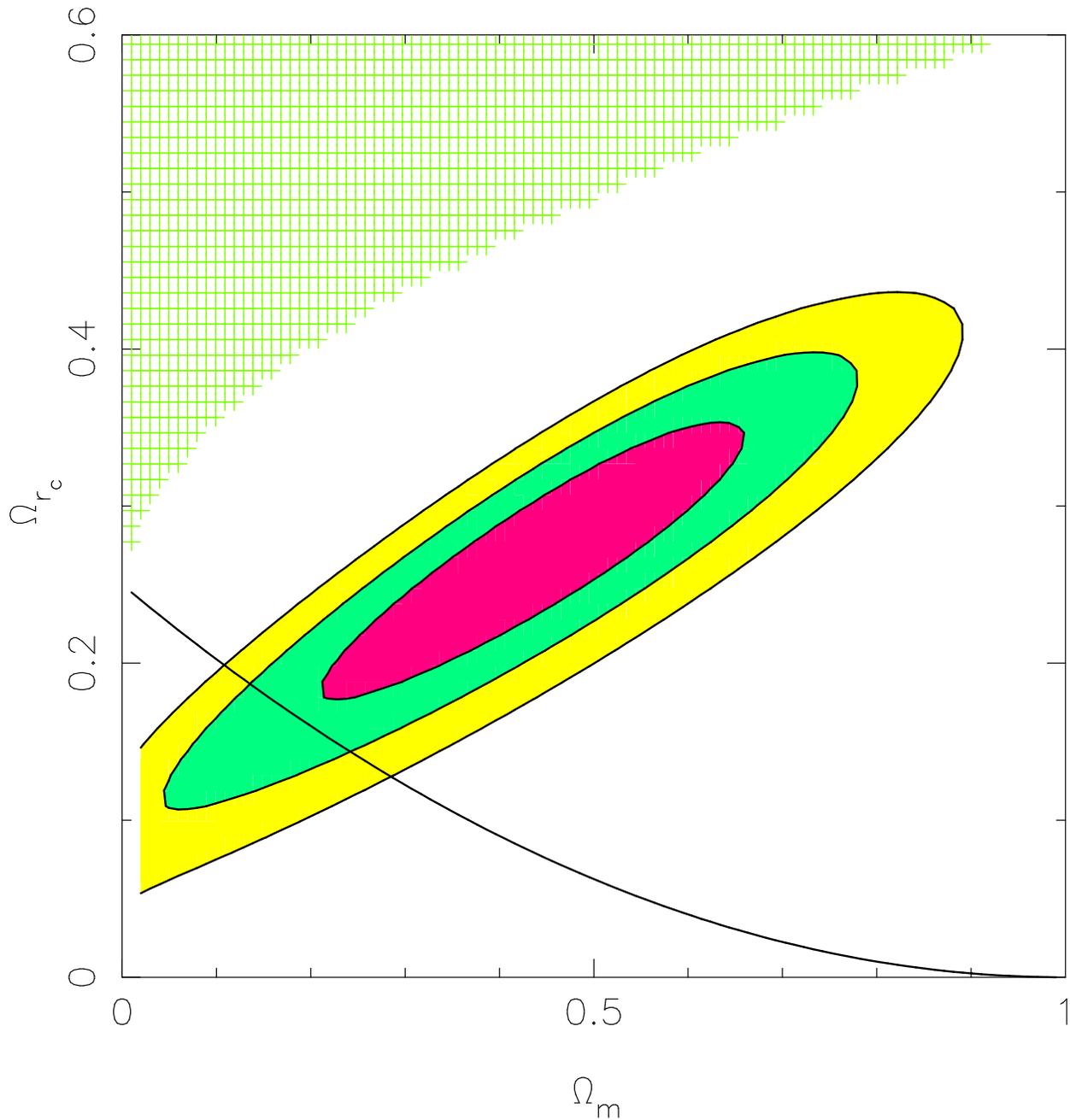}
\figcaption{Probability contours for $\Omega_{r_c}$ and $\Omega_m$ in the
	    model of gravitational leakage into an extra dimension, for the
	    172 SNeIa taken from Tonry et al. (2003) -- see the text for a
	    detailed description of the method.
	   The 68\%, 95\% and 99\% confidence levels in the $\Omega_{r_c}$ -
	    $\Omega_m$ plane are shown in red, green, and yellow shaded areas,
	    respectively.
	   The cross-hatched region at the upper left represents the 
	    ``no-big-bang'' region, while the thick solid line corresponds to 
	    the flat universe.
	   The best fit happens at $\Omega_m=0.43$ and $\Omega_{r_c}=0.26$.
	    \label{Fig_cont1} 
	   }
\end{figure}

\clearpage

\begin{figure}
\plotone{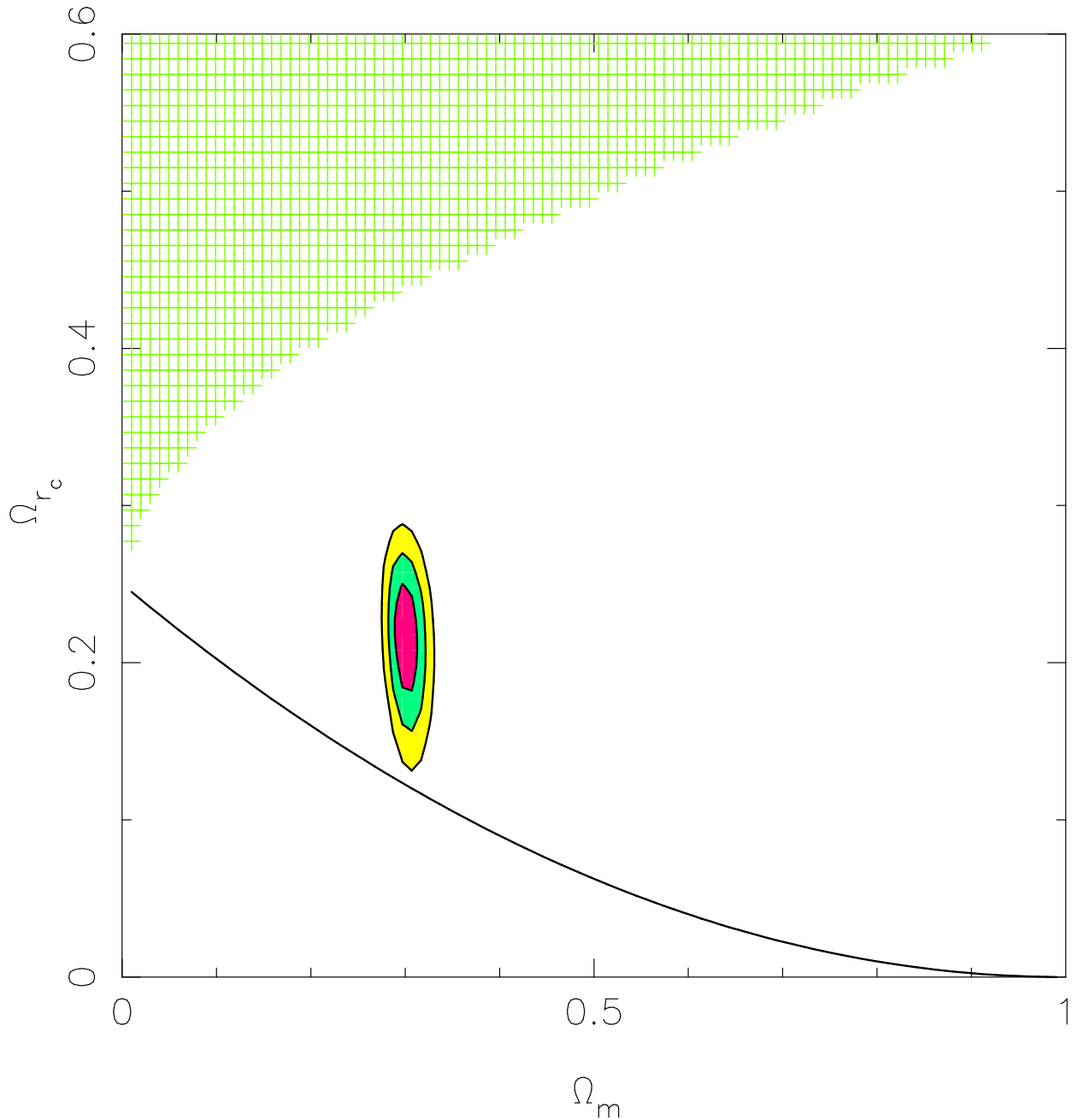}
\figcaption{Probability contours over $\Omega_{r_c}$ and $\Omega_m$ for the
	    combination of the 172 SNeIa taken from Tonry et al. (2003) and
	    the 9 X-ray clusters from Allen et al. (2002, 2003).
	   The 68\%, 95\% and 99\% confidence levels in the $\Omega_{r_c}$ -
	    $\Omega_m$ plane are shown in red, green, and yellow shaded areas,
	    respectively.
	   The cross-hatched region at the upper left represents the
	    ``no-big-bang'' region, while the thick solid line corresponds to 
	    the flat universe.
	   The best fit happens at $\Omega_m=0.29$ and $\Omega_{r_c}=0.21$,
	    hence giving a closed universe with $\Omega_k = -0.36$. 
	   However, the results depends on the X-ray gas mass fraction data
            from Allen et al. (2002, 2003), in which the errorbars might be
            on the optimistic side.
	    \label{Fig_cont12}
	   }
\end{figure}

\clearpage

\begin{figure}
\plotone{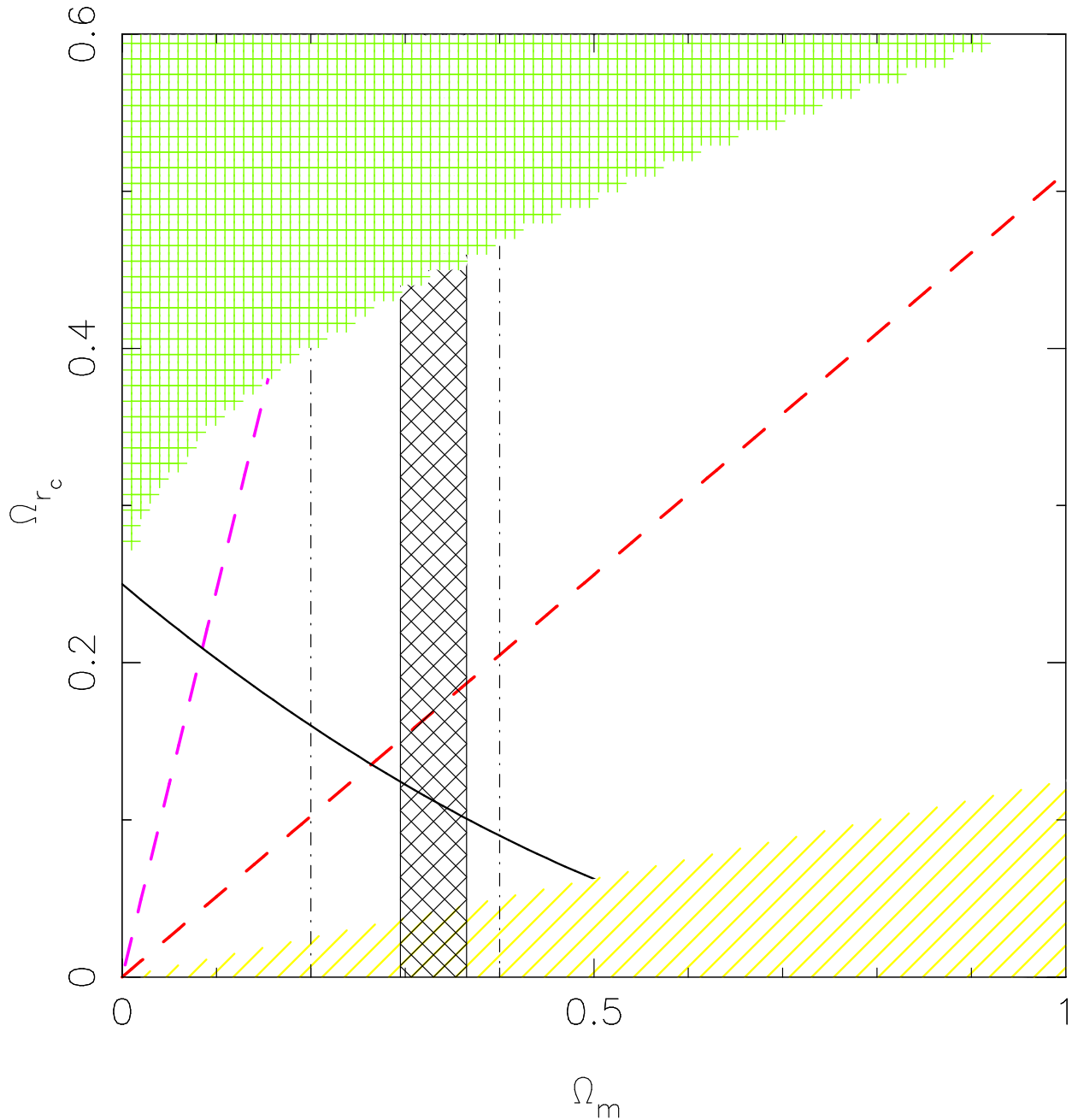}
\figcaption{Constraints on the parameter space ($\Omega_{r_c}$, $\Omega_m$)
	    of the model of gravitational leakage into an extra dimension
	    from the turnaround redshift of acceleration.
	   The hatched region at the lower right is the decelerating model,
	    and the cross-hatched region at the upper left is the closed
	    cosmological model without big bang.
	   The right and left dashed lines are $z_{q=0}=0.6, 1.7$, 
	    respectively, while the thick solid line is the flat universe.
	   Thus, in order to explain that acceleration happened earlier than
	    $z_{q=0}=0.6$, the gravitational leakage model needs a low
	    matter density, $\Omega_m < 0.27$, if the universe is flat.
	   The vertical strip with cross-hatching corresponds to the matter
	    density $\Omega_m=0.330\pm 0.035$ found by Turner (2002), which
	    clearly asks for a closed universe to explain $z_{q=0}>0.6$.
	   For convenience, we also draw two dot-dashed lines for 
	    $\Omega_m = 0.2, 0.4$, for which there are some ranges to be 
	    compatible with a flat universe.
	   We note that a matter density of $\Omega_m < 0.27$ is also
	    permitted by the WMAP data (Bennett et al. 2003).
	    \label{Fig_turnaround}
	   }
\end{figure}

\end{document}